\documentclass[journal,twocolumn,web]{ieeecolor}
\usepackage{generic}
\usepackage{amsmath,amssymb,amsfonts}
\usepackage{algorithmic}
\usepackage{graphicx}
\usepackage{textcomp}

\usepackage{booktabs}
\usepackage{subcaption}
\usepackage{multirow}
\usepackage[sorting=none, backend=biber, style=ieee, url=false, doi=false, maxbibnames=3]{biblatex}

\usepackage{worldflags}
\usepackage{tikz}
\usepackage{float}

\usepackage{array}
\usepackage{adjustbox}
\usepackage{hyperref}
\usepackage{dblfloatfix}
\newcommand{\supplementarysection}{%
  \setcounter{figure}{0}
  \let\oldthefigure\thefigure
  \renewcommand{\thefigure}{S\oldthefigure}
}

\def\BibTeX{{\rm B\kern-.05em{\sc i\kern-.025em b}\kern-.08em
    T\kern-.1667em\lower.7ex\hbox{E}\kern-.125emX}}
\begin{document}
%
\title{RawECGNet: Deep Learning Generalization for Atrial Fibrillation Detection from the Raw ECG}
\author{Noam Ben-Moshe, Kenta Tsutsui, Shany Biton,  Leif S{\"{o}}rnmo, \IEEEmembership{Fellow, IEEE}, \\ and Joachim A. Behar, \IEEEmembership{Senior Member, IEEE}
\thanks{Manuscript submitted on \today. Research was supported for NBM, SB, and JB by a grant (3-17550) from the Ministry of Science \& Technology, Israel \& Ministry of Europe and Foreign Affairs (MEAE) and the Ministry of Higher Education, Research and Innovation (MESRI) of France. SB, NBM, MS, and JAB acknowledge the support of the Technion-Rambam Initiative in Artificial Intelligence in Medicine, Hittman: Technion EVPR Fund: Hittman Family Fund and Israel PBC-VATAT and by the Technion Center for Machine Learning and Intelligent Systems (MLIS). (Corresponding author: jbehar@technion.ac.il)}
\thanks{N. Ben-Moshe, is with the Faculty of Computer Science and the Faculty of Biomedical Engineering, Technion-IIT, Haifa, Israel. S. Biton, and J. A. Behar are with the Faculty of Biomedical Engineering, Technion-IIT, Haifa, Israel. L. S\"ornmo is with the Department of Biomedical Engineering, Lund University, Lund, Sweden. K. Tsutsui is with the Department of Cardiovascular Medicine, Faculty of Medicine, Saitama Medical University International Medical Center, Saitama, Japan. Corresponding author: Joachim A. Behar (jbehar@technion.ac.il).}
}
\maketitle

\begin{abstract}

\textbf{Introduction}:  Deep learning models for detecting episodes of atrial fibrillation (AF) using rhythm information in long-term, ambulatory ECG recordings have shown high performance. However, the rhythm-based approach does not take advantage of the morphological information conveyed by the different ECG waveforms, particularly the f-waves. As a result, the performance of such models may be inherently limited. \textbf{Methods}: To address this limitation, we have developed a deep learning model, named RawECGNet, to detect episodes of AF and atrial flutter (AFl) using the raw, single-lead ECG. We compare the generalization performance of RawECGNet on two external data sets that account for distribution shifts in geography, ethnicity, and lead position. RawECGNet is further benchmarked against a state-of-the-art deep learning model, named ArNet2, which utilizes rhythm information as input. \textbf{Results}: Using RawECGNet, the results for the different leads in the external test sets in terms of the F1 score were 0.91--0.94 in RBDB and 0.93 in SHDB, compared to 0.89--0.91 in RBDB and 0.91 in SHDB for ArNet2. The results highlight RawECGNet as a high-performance, generalizable algorithm for detection of AF and AFl episodes, exploiting information on both rhythm and morphology.
\end{abstract}



\section{Introduction}
Atrial fibrillation (AF) is the most prevalent form of cardiac arrhythmia~\cite{ProspectiveMoti2015} and increases the risk of stroke fivefold~\cite{ESC2020Guidelines}. Automated analysis of long-term, ambulatory ECG recordings (Holter) has long been used to capture information on AF. Considerable research has been devoted to developing methods for AF detection, with the aim of increasing the detection accuracy and reducing the workload of clinical staff. AF detection has evolved from methods using rhythm features and morphological features, followed by thresholding or machine learning-based classification~\cite{SornmoAFbookDetection}, to arrive at today's large variety of deep learning-based detectors~\cite{Xiao2023}

To achieve high performance, it is crucial that deep learning models are robust, meaning that the models should generalize across distribution shifts related to, e.g., ethnicity, lead position, age, and sex. Indeed, although high performance has been reported for deep learning models using raw physiological signals~\cite{phan2019seqsleepnet,levy2022oxinet,ribeiro2020automatic}, the models tend to generalize poorly or moderately well on other data sets~\cite{kotzen2022sleepppg, levy2022oxinet, ballas2023towards}. 

Deep learning-based AF detectors designed to process single-lead ECGs can be broadly categorized according to whether the RR~intervals or the raw ECG serve as input to the network, with increasing attention paid to the latter category. For example, in Mousavi et al.~\cite{Han2020}, the raw ECG was used as input to a method composed of three parts, each with a stacked bidirectional recurrent neural network, followed by an attention layer; however, generalization to external data sets was not demonstrated. In Kr{\'{o}}l-J{\'{o}}zaga~\cite{AF2D2022}, the raw ECG was transformed into 2D images which were classified with respect to the presence of AF. Although this approach achieved good results on a local test set, not all available leads were processed or the generalizability of the model established. Liu et al.~\cite{Liu2024}, presented an alternative approach to AF detection, using the raw ECG data as input to the network and the RR~intervals as input to a post-processing step. However, low generalization performance was reported when their model was applied to external data sets. The potential benefit of employing transfer learning was demonstrated, specifically using a small sample from the target domain data sets, to enhance the generalizability of the model. However, to obtain an annotated subset for each possible target domain is highly impractical. 

A number of deep learning-based detectors have been designed to find episodes of both AF and atrial flutter (AFl) using the same network \cite{Carrara2015, chocron2020remote,RawAFAFLDetection2021, Biton2023}. In Wang et al.~\cite{RawAFAFLDetection2021}, the raw ECG was inputted to a modified bidirectional, long short-term memory network, however, generalization to external data sets was not investigated. In Biton et al.~\cite{Biton2023}, the RR~intervals served as input to residual network blocks (ResNet) and gated recurrent units (GRU).  The detector demonstrated high performance and strong generalizability, however, most AFl episodes were missed. This was because AFl is a regular rhythm that is easily confused with normal sinus rhythm.

In a study comparing the performance evaluation of 14 different deep learning-based AF detectors, all using the MIT--BIH AF database for testing, only one detector was trained on a database different from the test database~\cite{Butkuviene2021AFdetection}. Additionally, 10 of these detectors were trained and cross-validated on data sets that included the same patients, a type of information leakage that leads to inflated results. When comparing the performance of the same network models, measures such as F1, sensitivity, and positive predictive value were found to be approximately 15\% lower in the absence of information leakage than in the presence~\cite{Xiao2023}. The work by Zou et al.~\cite{Zou2023} is one of the very few that validates a deep learning-based AF detector on four external data sets. This work uses a raw, single-lead ECG as input and has the merit of evaluating generalization on independent target domains. However, it did not evaluate performance across different leads, age, sex, and the test data sets were small in terms of the number of patients and recording length.  Thus, the capability of different deep learning-based AF detectors to generalize across different population samples deserves to be given further attention. 

To our knowledge, very limited research has rigorously compared the performance of RR~interval and raw ECG-based, single-lead AF detectors. No previous research has evaluated the generalization performance of raw ECG-based AF detectors across important sources of distribution shift including the lead, geography, sex and age. There is a pressing need for the development of a robust model that exhibits a high degree of generalization to ensure its efficacy when deployed in a variety of health settings and population samples~\cite{Behar2023}.

This paper starts by introducing the architecture and strategy used to train RawECGNet and create a generalizable representation of the raw underlying ECG data for AF detection. Then, a set of experiments is then proposed to evaluate the performance of the model in data sets across distribution shifts. Results are provided on the source domain test set and two independent target domain data sets, totaling an overall of 7,404 hours of manually annotated test data. This research makes the following key contributions: 
\begin{itemize}
    \item RawECGNet, a new deep learning model for AF detection based on the raw, single-lead ECG; 
    \item Rigorous benchmarking of RawECGNet against a state-of-the-art, rhythm information-based model ArNet2; 
    \item Evaluation of generalization performance on two independent data sets from different geographical locations as well as across lead position, age, and sex; 
    \item A thorough quantitative error analysis pinpointing the main sources of false positives and false negatives.
\end{itemize}

\smallskip\noindent
\begin{small}
\begin{tabular}{l}
\end{tabular}
\end{small}

\section{Materials}
\subsection{Data sets}
Within this framework, some data sets are used for model development (source domain) while other data sets are used for testing purposes (target domain). Three different data sets were used: The University of Virginia AF data set (UVAF), USA \cite{Carrara2015,Chugh2014WorldwideStudy}, the Rambam Hospital Holter clinic data set (RBDB), Israel \cite{Biton2023}, and the Saitama Hospital data set (SHDB), Japan \cite{Biton2023}. The RBDB and SHDB were used to test generalization. All test set recordings were manually annotated at the beat level~\cite{Biton2023} and a description of the test sets is shown in table \ref{table:DemographicDescription}. Data usage is approved by Rambam Health Care Campus Institutional Ethics Committee (IRB D-0402-21) and Saitama Medical University Institutional Ethics Committee (IRB 20–030).

UVAF consists of three-lead Holter recordings for which no lead label information is available. The original data set consists of 2,242 patients totaling 51,386 hours of continuous ECG recordings. As no patient reports were available, diagnoses were inferred from the annotation of AF episodes~\cite{Biton2023}. UVAF was divided into a training set ($n = 2137$) and a test set ($n = 105$). The recordings were obtained using Philips Holter software and digitized at a rate of 200 Hz. For most recordings, three leads were available but without any indication of the lead system used. Each beat was automatically annotated with regard to rhythm type, and about half of the annotations of all recordings were reviewed by medical school students. 

RBDB includes medical reports and three-lead Holter recordings (leads CM5, CC5, and CM5R), except when battery life had to be saved to handle recordings exceeding 24~h, then resulting in two-lead recordings (CM5 and CC5). This data set ($n = 116$) was recorded using the PathFinder Holter monitor and digitized at a rate of 128 Hz. 

SHDB consists of medical reports and two-lead Holter ECG recordings ($n = 100$, leads NASA and CC5), acquired using the Fukuda Holter monitor and digitized at a rate of 125 Hz.

The combined test set is made up of all the leads from the three test sets.

\begin{table}
\caption{Description of the test sets. Age is presented as median and interquartile range (Q1--Q3).}
\label{table:DemographicDescription}
\centering
\begin{tabular}{lccl}
\toprule
& \textbf{UVAF}& \textbf{SHDB}& \textbf{RBDB}\\
       \midrule
    Origin & USA & Japan & Israel \\[0.5ex] 
    Patients, $n$ & 105 & 100 &116 \\  [0.5ex]
    Age (yrs) & 69 (59--77) & 70 (62--75) & 70 (58--78) \\  [0.5ex]
    Female, $n_f$ & 51 & 45 & 55 \\  [0.5ex]
    AF prevalence\% & 45 & 33 & 19  \\  [0.5ex]
    AFl prevalence \% & 2 & 4&2 \\  [0.5ex]
\bottomrule
\end{tabular}
\end{table}

\section{Methods}

\subsection{Preprocessing} \label{secPreprocess}
Each recording is divided into 30-s non-overlapping windows. For each window, the rhythm label is defined per the rhythm most represented over the window. Similarly to~\cite{Moss2014, RawAFAFLDetection2021, chocron2020remote, Biton2023}, a combined label of AF and AFl is introduced, denoted AF\textsubscript{l}. Noisy windows are excluded from the training set using the signal quality index bSQI, defined by the fraction of beats detected by one QRS detector matching those detected by another detector~\cite{Behar2013ECGReduction, Li2008RobustSources}; the exclusion threshold is set to 0.8. All recordings are resampled to 200~Hz. A recording is excluded from the study whenever one of the following three exclusion criteria is fulfilled: Patients under 18 years of age, missing beat labels, and more than 75\% of the windows labeled as poor quality, cf.~\cite{chocron2020remote}.


\subsection{RawECGNet architecture}
\begin{figure}[t!]
\centering
\includegraphics[scale= 0.7]{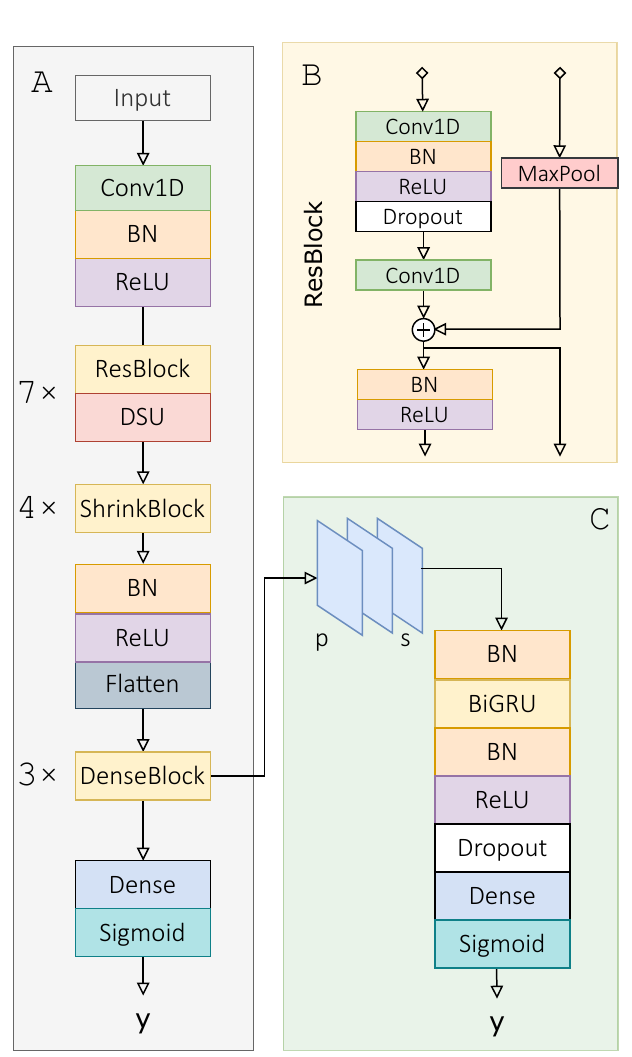}
\caption{RawECGNet architecture. (a)~Network for training (first step), consisting of residual network blocks (ResBlocks), shrink blocks, batch normalization (BN), a domain-shifts-with-uncertainty (DSU) layer, and dense blocks. (b)~The ResBlock architecture. (c)~The second step consists of a bidirectional GRU (BiGRU) unit and a dense block. The first step of the network is trained for binary classiﬁcation of 30-s windows, and the extracted features of each window are used as input to the second step. Each window is concatenated with $p$ preceding and $s$ succeeding windows. A dense layer is used and outputs for each window a probability of AF\textsubscript{l}.}
\label{model_architecture}
\end{figure}

The model takes as input 30-s windows of the raw ECG and produces a binary output, AF\textsubscript{l} or non-AF\textsubscript{l}. The model consists of 2.5 million parameters. The training process is divided into two distinct steps: The first involves an encoder that is tasked with learning a representation of the underlying ECG signal. In the second step, a recurrent neural network is employed, which facilitates the accommodation of the time-dependent characteristics inherent in the lengthy and continuous time series. Fig. ~\ref{model_architecture} displays a schematic block diagram of the RawECGNet architecture. 

In the first step, the ResNet architecture~\cite{ResNetOrig} is used to extract features. Each ResNet block consists of 1D-CNN layers with batch normalization layers followed by a rectified linear unit (ReLU) activation function and a shortcut connection with max pooling. A domain-shifts-with-uncertainty (DSU) layer is added after each ResNet block~\cite{DSULi2022}, whose structure is based on the assumption that the statistics of the extracted features follow a multi\-variate Gaussian distribution. The DSU layer handles situations when the test domains bring statistical shifts with different potential directions and intensities compared to the training domain. The DSU layer models each batch feature statistics as a random variable, computes its empiric mean and variance, and, with a certain probability, resamples the statistics of the features. It is assumed that by modeling the uncertainty of the feature statistics while training, the model will gain better robustness against potential distribution shifts from the source domain and will offer better generalization. The DSU layer is applied with a probability of~0.5. Shrink blocks have a structure similar to that of ResNet blocks, except that the number of channels decreases by a factor of two rather than increases by a factor of two between neighboring blocks. Shrinking is introduced to shorten the dimension of the feature space. Dense blocks have a structure defined by a dense layer, batch normalization, leaky rectified linear unit (LReLU) activation, and dropout regularization. The architecture consists of seven ResNet blocks followed by four shrink blocks and three dense blocks. Between each dense block, the hidden dimension is divided by two.

In the second step of the training process, the hidden dimension is computed for each window. To provide contextual and temporal information, the current window is combined with enclosing windows and then fed to a bidirectional gated recurrent unit (BiGRU) layer, which is well suited for handling temporal data. Subsequently, the output of the BiGRU layer undergoes a series of transformations, including batch normalization, LReLU activation, dropout regularization, and, finally, a dense layer. The last dense layer, made up of a single unit, generates the binary output assigned to the current window.

In both training steps, the loss function is given by a weighted binary cross-entropy which assigns more weight to AF\textsubscript{l} windows to account for the fact that the two classes AF\textsubscript{l} and non-AF\textsubscript{l} are highly imbalanced. 

It should be noted that all three leads of the recordings in the training set and the validation set are used to prevent the model from overfitting to a specific lead representation. The threshold on the output probabilities is chosen as the point that maximizes F1, defined in Section~\ref{secPerformance}, in the validation set. 

\subsection{Benchmark}
RawECGNet is benchmarked against a state-of-the-art deep learning model, named ArNet2 \cite{Biton2023}, using RR~intervals as input. In previous work, we demonstrated the superiority of ArNet2 against other algorithms for AF detection using RR~intervals, including ArNet~\cite{chocron2020remote}, the AFEv-based algorithm~\cite{Sarkar2008}, and a gradient boosting (XGB) model using heart rate variability  features~\cite{Behar2018PhysioZoo}. 

\subsection{Performance measures} \label{secPerformance}
The following performance measures are used to evaluate the performance of the model for the classification of AF: Sensitivity (Se), specificity (Sp), positive predictive value (PPV), area under the receiver operating characteristic (AUROC), and harmonic mean between Se and PPV (F1). The performance measures are computed by grouping AF and AFl windows together as the positive class as well as for the positive class defined by AF windows or AFl windows exclusively. The confidence interval is estimated for F1 using bootstrapping. Each test set is bootstrapped 100 times. The nonparametric Mann--Whitney rank test is used to determine whether the models are significantly different for~F1, where $p < 0.05$ indicates a statistically significant difference.

RawECGNet takes 30-s windows as input, whereas ArNet2 takes 60-beat windows as input. To properly benchmark the networks, their outputs are aligned to provide binary classification, i.e., AF\textsubscript{l} or $\textrm{non-AF}_{\textrm{l}}$, for each 5-s window. For 5-s windows with more than one model's input window overlapping, the prediction of the window containing the greater part of the 5-s window is used as prediction for that window. In this way, the same windows are evaluated by both algorithms, exemplified in Fig. S3. The 5-s length is selected as a compromise between longer windows, which may result in windows with a mix of rhythms, and shorter windows, which may not cover several heartbeats. In the test set, no windows are excluded. Performance is measured for each lead in each data set.

Another part of the performance evaluation relates to AF burden, defined by~\cite{chocron2020remote}:
\begin{equation}
B_{\textrm{AF}} = \frac{\displaystyle\sum_{i=1}^{N} l_{i} \cdot I_{i}}{\displaystyle\sum_{i=1}^{N} l_{i}},
\label{AFB_eq}
\end{equation}
where $l_{i}$ is the length of the $i$:th window, $I_{i}$ is equal to 1 for AF\textsubscript{l} and 0 otherwise, and $N$ is the total number of windows. 
The estimation error of $B_{\textrm{AF}}$ is defined by the following percentage~\cite{chocron2020remote}:
\begin{equation}
    E\textsubscript{AF} = \frac{\displaystyle\sum_{i=1}^{N} l_{i} \cdot (\hat{y}_{i}-y_{i})}{\displaystyle\sum_{i=1}^{N} l_{i}} \cdot 100 \quad (\%),
\label{EAF_eq}
\end{equation}
where $y_{i}$ is a binary value representing the window label and $\hat{y}_i$ is the binary value predicted by the model, equal to 1 for AF\textsubscript{l} and 0 otherwise. The $E$\textsubscript{AF} is computed for different AF burdens, defined by the intervals $[0,0.04]$, $(0.04,0.8]$, and $(0.8,1]$, which in the following are referred to as mild AF, moderate AF, and severe AF, respectively~\cite{chocron2020remote}. In addition to these three groups of AF burden, the fourth group is non-AF.

\subsection{Ablation study}
The main contributors to the detection performance and the generalization capability of RawECGNet are determined by systematically inputting key components of the model, namely the use of a single lead versus multiple leads for model training, the use of the BiGRU layer and the use of the DSU layers. The effect of each component on F1 is shown in Fig.~\ref{ablation_study}.

\subsection{Error analysis}
The false negative (FN) and false positive (FP) windows are quantitatively investigated over the combined test set using the following five criteria: 1)~poor signal quality, see definition in~Sec.~\ref{secPreprocess}, 2)~whether FN is AFl or AF, 3)~mixed labels, defined as a mixture of AF\textsubscript{l} and normal sinus rhythm, 4)~atrial tachycardia (AT) or atrial bigeminy (AB), and 5)~a heart rate exceeding 100~bpm. For the remaining FN and FP, 200 FP and 200 FN were provided to an electrophysiology expert (co-author KT) for manual inspection and labeling of the following categories: 6)~incorrect reference label, 7)~presence of ectopic beats 8)~other.  This enabled us to further estimate the percentages of examples falling in these categories.

\section{Results}

\subsection{Per window classification}
\begin{figure}[!ht]
\centering
\includegraphics[scale=0.6]{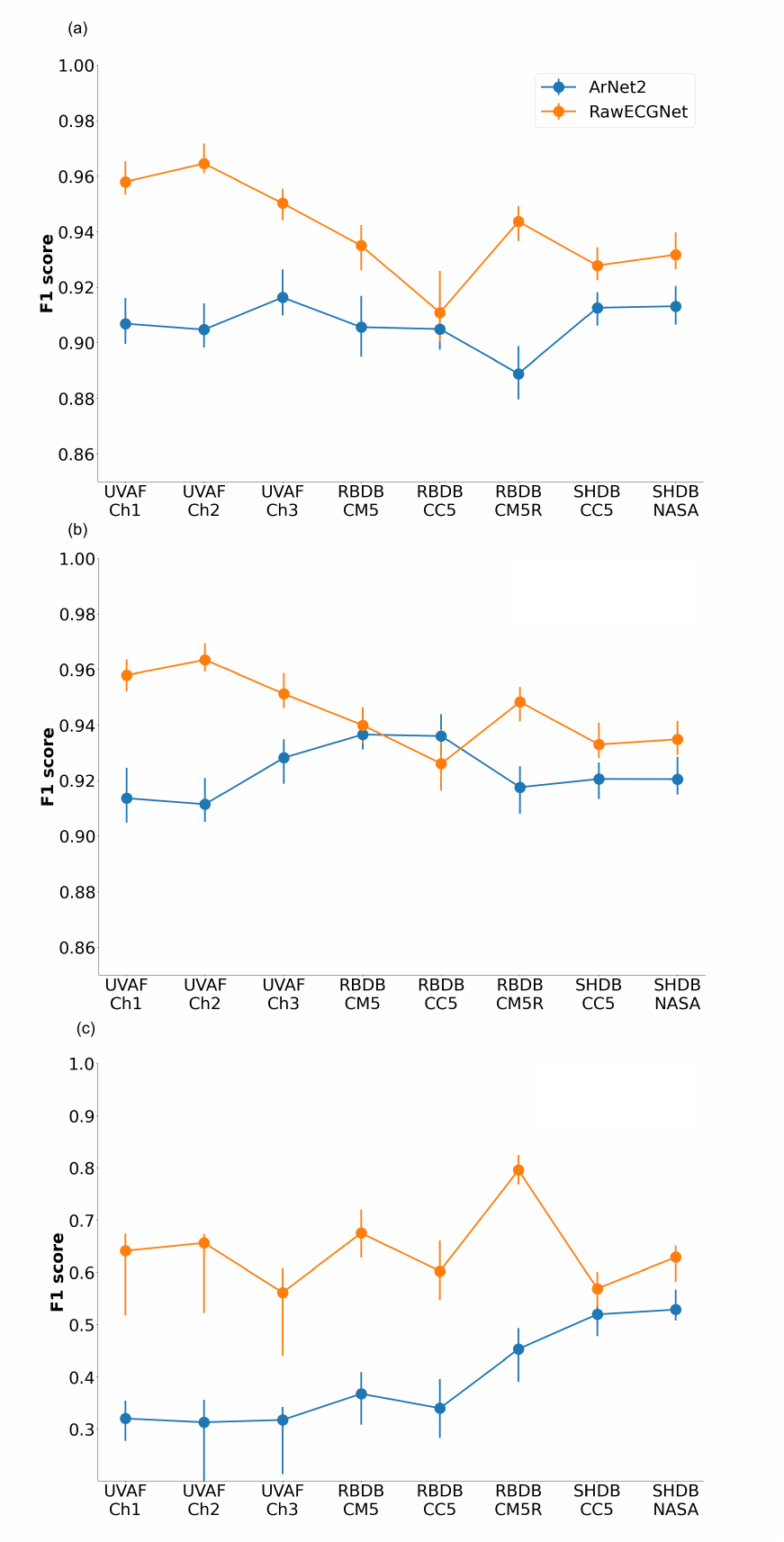}
\caption{Performance of AF\textsubscript{l} window classification for ArNet2 and RawECGNet, presented for each lead and for each data set. The median F1 bootstrap results are presented with error bars representing Q1 and Q3 of the results. The following windows are analyzed: (a)~All windows, (b)~AF and non-AF\textsubscript{l} windows only, and (c) AFl and non-AF\textsubscript{l} windows only.}
\label{f1_results_all}
\end{figure}

The performance in terms of F1 on the source domain test set (UVAF) is found to be in the range of 0.95--0.96 for RawECGNet, compared to 0.90--0.91 for ArNet2.  The ranges signify the variability in F1 among different leads, with the lowest and highest scores presented. In those cases where no range is indicated, all leads achieve identical scores. The performance on the two external test sets, modeling shifts in the distribution of geography and leads, is 0.91--0.94 (RBDB) and 0.93 (SHDB), compared to 0.89--0.91 and 0.91, respectively, for ArNet2. The performance statistics are shown in Table~\ref{tab:generalization_resutls} and Fig.~\ref{f1_results_all}(a).

Excluding AFl windows from the test sets and considering AF versus non-AF classification, the resulting F1 is 0.95--0.96 for RawECGNet and 0.91--0.93 for ArNet2 on UVAF, 0.92--0.95 for RawECGNet and 0.92--0.94 for ArNet2 on RBDB, and 0.93--0.94 for RawECGNet and 0.92 for ArNet2 on SHDB, see Fig. \ref{f1_results_all}(b).
In RBDB, leads CM5 and CM5R are associated with better performance for RawECGNet than ArNet2, whereas the reverse applies to lead CC5. 

When AF windows are excluded from the test sets so that the capability of RawECGNet to detect AFl episodes can be analyzed, RawECGNet outperforms ArNet2 significantly for the source domain test set as well as the two target domains, see Fig.~\ref{f1_results_all}(c).

\begin{figure}[!ht]
\centering
\includegraphics[scale=0.6]{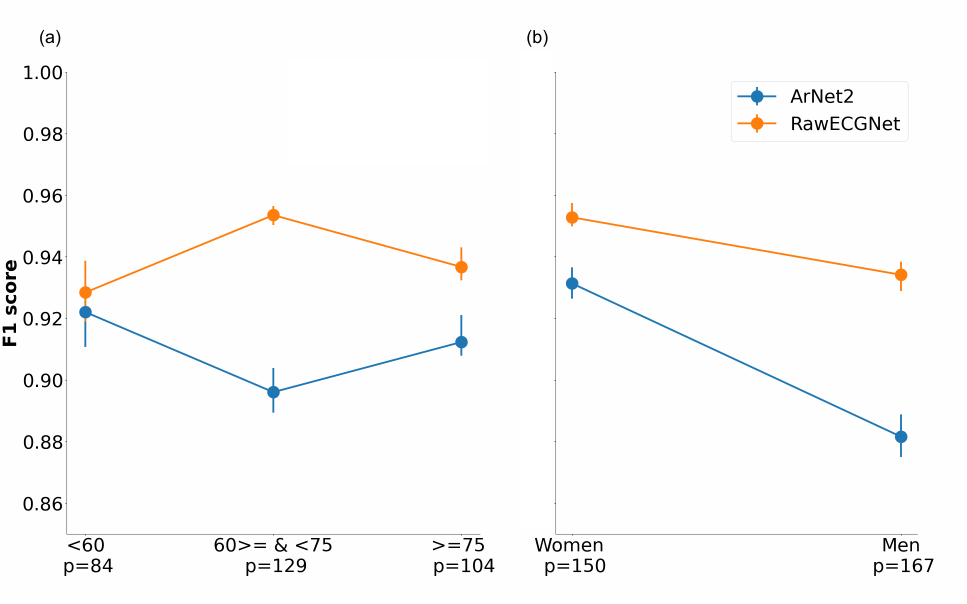}
\caption{Performance of AF\textsubscript{l} window classification across (a)~age and (b)~sex. The median F1 bootstrap results are presented with error bars representing Q1 and Q3 of the results. The results are presented for the combined test set. The number of patients $(p)$ is displayed for each group.}
\label{age_sex_comparison}
\end{figure}

For the combined test set, Fig.~\ref{age_sex_comparison} shows F1 for men and women and for different age groups. RawECGNet performed significantly better than ArNet2 for all age groups and sex.

\begin{table*}
    \centering
    \caption{Performance statistics of the models across different geographical distributions for the task of  AF\textsubscript{l} windows classification. The UVAF is the source domain while RBDB and SHDB are target domains. In $|$E\textsubscript{AF}(\%)$|$ $\mu$ represents the median value.}
    \begin{tabular}{lcccccccc}
          & & & & \centering F1 & \centering AUROC & \centering Se & \centering  Sp & $|$E\textsubscript{AF}(\%)$|$ $\mu$ (Q1-Q3)\\[0.5ex] 
         \midrule
         \multirow{6}{*}{\rotatebox[origin=c]{90}{\parbox[c]{1cm}{\centering Source}}} & \multirow{6}{*}{\rotatebox[origin=c]{90}{\parbox[c]{1cm}{\centering UVAF \worldflag[length=4mm, width=2.5mm]{US}}}} & \multirow{2}{*}{\centering Ch1}  & \centering ArNet2 & 0.91 & 0.97 & 0.87& \textbf{0.96} & 0.09 (0.0-3.29)\\  
        & & & \centering RawECGNet & \textbf{0.96} & \textbf{0.99} &\textbf{0.96} & \textbf{0.96} & \textbf{0.02 (0.0-0.99)} \\ \cline{3-9}

        &  &\multirow{2}{*}{\centering  Ch2} & \centering ArNet2 & 0.91 & 0.97 & 0.87& 0.95 & 0.09 (0.0-3.29)\\  
        & & & \centering RawECGNet & \textbf{0.97} & \textbf{0.99} &\textbf{0.97} & \textbf{0.96} & \textbf{0.00 (0.0-0.60)} \\ \cline{3-9}

        & &\multirow{2}{*}{\centering Ch3}  & \centering ArNet2 & 0.92 & 0.97 & 0.89& \textbf{0.95} & 0.02 (0.0-3.14)\\  
        & & & \centering RawECGNet & \textbf{0.95} & \textbf{0.98} &\textbf{0.96} & \textbf{0.95} & \textbf{0.00 (0.0-1.19)} \\ 
        \midrule
        
        \multirow{12}{*}{\rotatebox[origin=c]{90}{\parbox[c]{1cm}{\centering Target}}}&\multirow{6}{*}{\rotatebox[origin=c]{90}{\parbox[c]{1cm}{\centering RBDB \worldflag[length=4mm, width=2.5mm]{IL}}}} &\multirow{2}{*}{\centering CM5}  & \centering ArNet2 & 0.91 & 0.98 & 0.88& \textbf{0.96} & 0.17 (0.0-4.54)\\  
        & & & \centering RawECGNet & \textbf{0.93} & \textbf{0.99} &\textbf{0.92} & \textbf{0.96} & \textbf{0.14 (0.0-1.24)} \\ \cline{3-9}

        & &\multirow{2}{*}{\centering CC5}   & \centering ArNet2 & \textbf{0.91} & \textbf{0.98} & \textbf{0.88}& 0.96 & 0.17 (0.0-4.44)\\  
        & & & \centering RawECGNet & \textbf{0.91} & \textbf{0.98} &\textbf{0.88} & \textbf{0.97} & \textbf{0.14 (0.0-0.86)} \\ \cline{3-9}

        & &\multirow{2}{*}{\centering CM5R} & \centering ArNet2 & 0.89 & 0.98 & 0.87& 0.94 & 0.46 (0.0-7.57)\\  
        & & & \centering RawECGNet & \textbf{0.94} & \textbf{0.99} &\textbf{0.92} & \textbf{0.97} & \textbf{0.14 (0.0-0.65)} \\ \cline{2-9}

         &\multirow{4}{*}{\rotatebox[origin=c]{90}{\parbox[c]{1cm}{\centering SHDB \worldflag[length=4mm, width=2.5mm]{JP}}}} & \multirow{2}{*}{\centering  CC5}  & \centering ArNet2 & 0.91 & \textbf{0.99} & 0.91& \textbf{0.98} & 0.34 (0.02-2.34)\\  
        & & & \centering RawECGNet & \textbf{0.93} & \textbf{0.99} &\textbf{0.93} & \textbf{0.98} & \textbf{0.31 (0.07-1.05)} \\ \cline{3-9}

        & &\multirow{2}{*}{\centering NASA}  & \centering ArNet2 & 0.91 & \textbf{0.99} & 0.91& \textbf{0.98} & 0.36 (0.01-2.34)\\  
        & & & \centering RawECGNet & \textbf{0.93} & \textbf{0.99} &\textbf{0.93} & \textbf{0.98} & \textbf{0.27 (0.04-0.90)} \\ 
         \bottomrule
    \end{tabular}
    \label{tab:generalization_resutls}
\end{table*}

\subsection{AF burden estimation}
\begin{figure}[!ht]
\centering
\includegraphics[page=1,width=1\columnwidth]{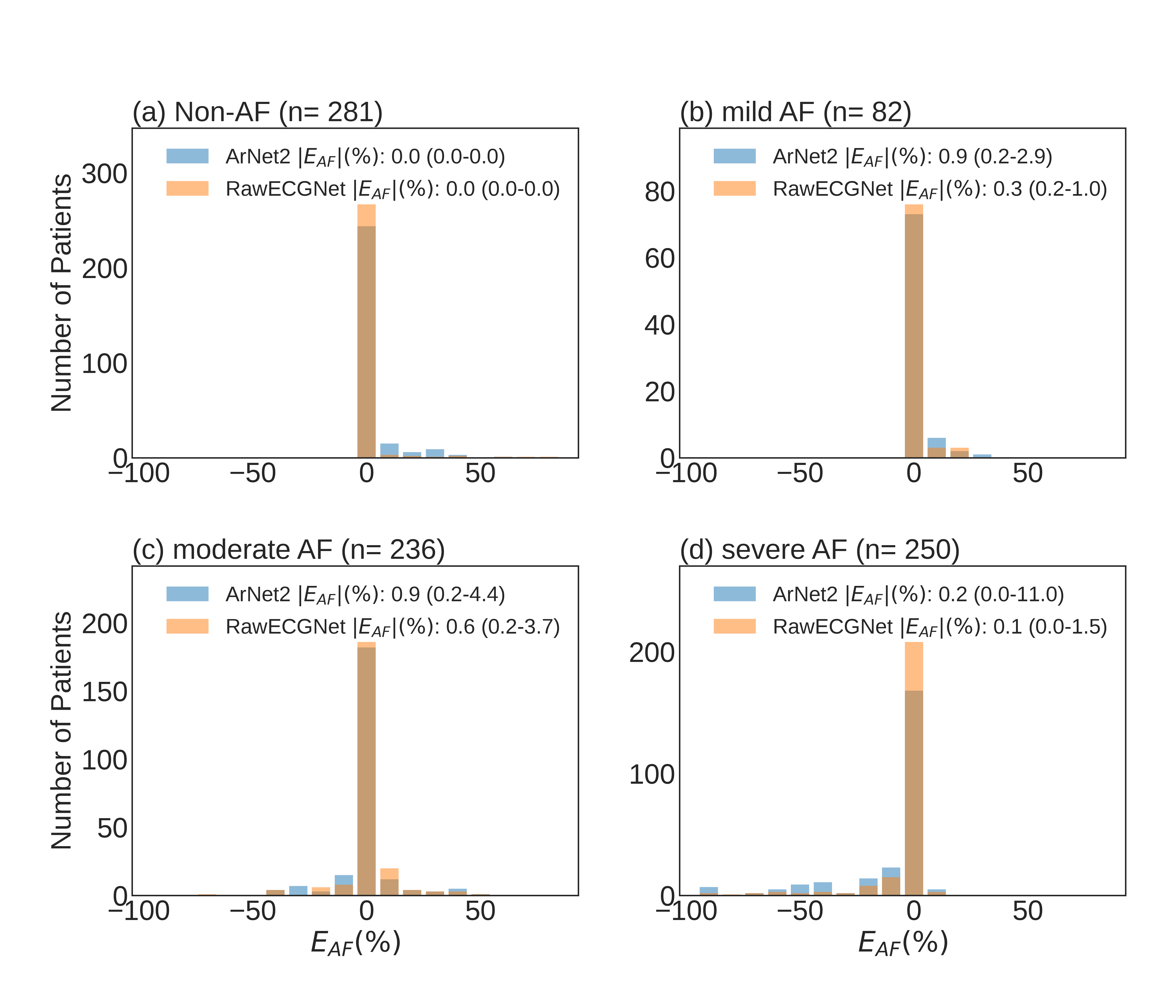}
\caption{Histograms of the AFB error $E\textsubscript{AF}$ (\%), computed for the combined test sets, including all leads, and grouped according to the following labels: (a)~Non-AF, (b)~mild AF, (c)~moderate AF, and (d)~severe AF.}
\label{all_error_sevirity}
\end{figure}

For the combined test set, Fig.~\ref{all_error_sevirity} presents $E\textsubscript{AF}$ for each group of AF burden and each model. For all groups, RawECGNet offers a performance superior to that of ArNet2. The improvement is particularly striking for the group with severe AF burden: RawECGNet produces a Q3 error of 1.5, while ArNet2 a Q3 error of~11.0. Table~\ref{tab:generalization_resutls} provides an overview of $E\textsubscript{AF}$ for each lead within each data set.  RawECGNet exhibits the lowest error in all leads.

\subsection{Ablation Study}
\begin{figure}[!ht]
\centering
\includegraphics[page=1,width=1\columnwidth]{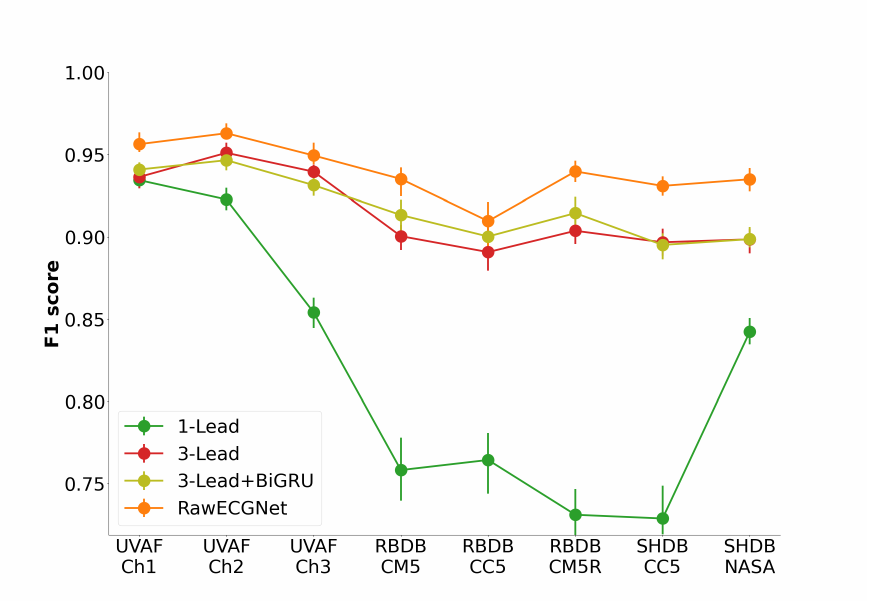}
\caption{Ablation study of the different components of RawECGNet.}
\label{ablation_study}
\end{figure}

When training the model using only one ECG lead and excluding the DSU layer and BiGRU, the resulting model, named the 1-Lead model, leads to poor generalization performance, see Fig.~\ref{ablation_study}. In contrast, when leveraging the three leads for training, the resulting model generalized significantly better on the two external test sets. This indicates that using a single lead for model training severely limits the capability of the model to generalize to different lead placements, while training on multiple leads avoids overfitting to a specific lead and improves generalization. When the BiGRU layer is added, the performance improved slightly, see Fig.~\ref{ablation_study} and the 3-Lead+BiGRU model. Finally, the inclusion of the DSU layer significantly improves generalization performance, see Fig.~\ref{ablation_study} and RawECGNet. These findings underscore the critical role of using multiple leads for training as well as the DSU layer. 

\begin{figure}[hbtp]
\centering
\includegraphics[width=0.8\columnwidth]{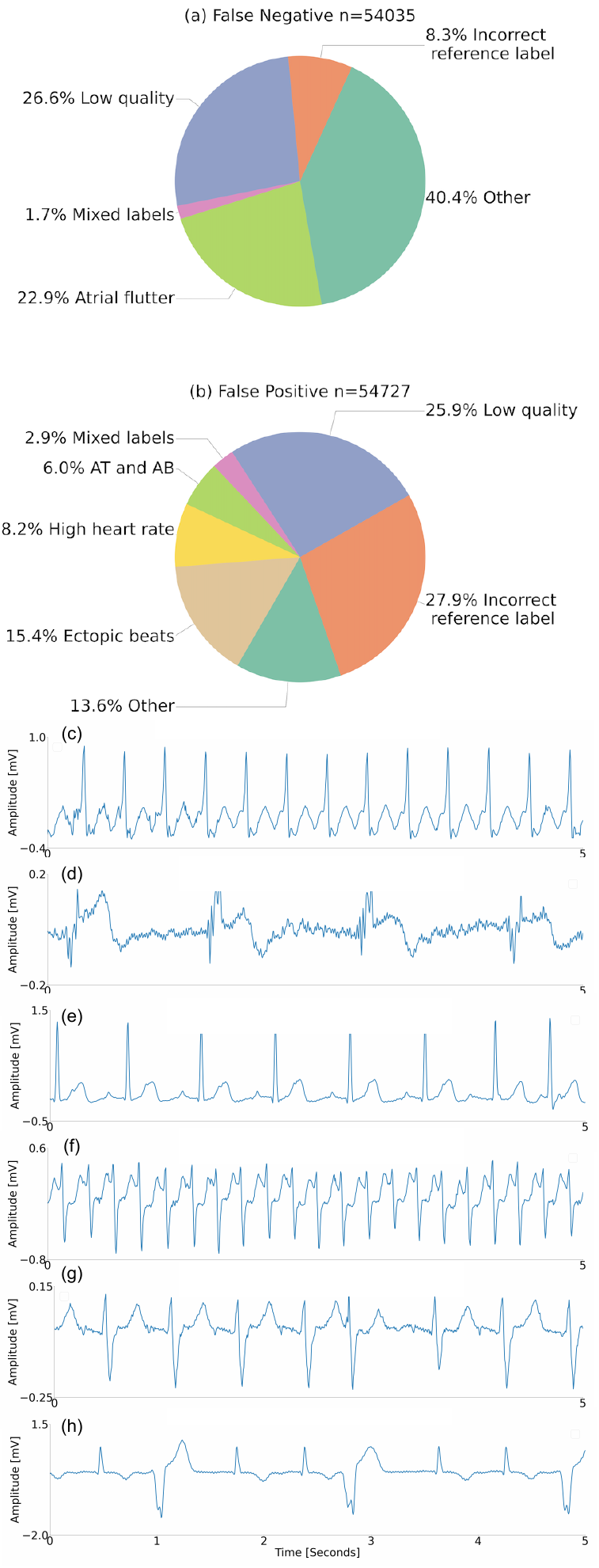}
\caption{Pie chart describing the repartition of false positives (FPs) and false negatives (FNs) across the three test sets. (a)~False negatives and (b)~False positives. The number of windows (n). Quantitative results are provided on overall FN and FP for the categories: low quality, mixed labels, atrial flutter, high heart rate, and other arrhythmias. Estimated percentages obtained by expert manual inspection of 200 FP and 200 FN are provided for the categories: incorrect reference label and ectopic beats. All examples that could not be interpreted with enough certainty are labeled as Other. (c)~An FN window with AFl. (d)~A low-quality FN window. (e)~An FP window with atrial tachycardia. (f)~An FP window with a high heart rate.  (g)~An FP window with mixed labels. Normal sinus rhythm is present in the early part of the window, and AF in the late part. (h)~An FP window with ectopic beats.}
\label{error_analysis_pies}
\end{figure} 

\subsection{Error Analysis}
In the combined test set, 27\% of all FNs are of low quality (bSQI less than 0.8). Of the remaining FN windows, 23\% are labeled AFl and 2\% have mixed labels. 26\% of all FPs are of low quality (bSQI less than 0.8). Among the remaining FP windows, 6\%  are labeled as AT and AB, 8\% have a heart rate greater than 100~bpm, and 2\% have mixed labels. From the expert inspection of  200 FN it is estimated that  8\% have an incorrect reference label. From the expert inspection of  200 FP it is estimated that 28\% have an incorrect reference label and 15\% have ectopic beats. A pie chart that summarizes the error analysis is shown in Fig.~\ref{error_analysis_pies}.


\section{Discussion}  
The main contribution of this research is the development of a robust, high-performance, generalizable, deep learning model for detecting AF episodes in single-lead ECGs. RawECGNet significantly outperformed ArNet2 for all individual leads of the local test set as well as the two external data sets. Moreover, RawECGNet reached a lower $E\textsubscript{AF}$ than ArNet2 as seen in Fig.~\ref{all_error_sevirity}.

The AF detection performance of RawECGNet was better than that of ArNet2, except for lead CC5, whereas AFl detection performance was considerably better for all leads. Ben-Moshe et al. \cite{BenMoshefwaves2023} showed that CM5R in RBDB and NASA in SHDB produced the best performance for f-wave extraction, thanks to their proximity to the atria. These observations are consistent with the results obtained by RawECGNet, which performed the best in detecting AF in these leads. The worst performance was obtained for lead CC5 in RBDB, which was also the worst performing lead for f-wave extraction~\cite{BenMoshefwaves2023}. 

Although RawECGNet offers better detection of AFl, error analysis showed that a large percentage of FN windows contained AFl. This may be explained by the fact that the percentage of AFl windows in the training set was very small (0.43\%), as opposed to the percentage of AF windows (6.71\%) in the training set. In this respect, increasing the percentage of AFl windows may yield a better representation of the AFl windows. In both RBDB and SHDB, many more AFl windows were missed in lead CC5 than in the other leads of the same data set. This result is expected since lead CC5 is far away from the right atria, thereby causing the atrial activity to be low in amplitude. Thus, for single-lead detection, the results show that lead placement has to be chosen close to the right atrium. Fig. S1 shows an example of an AF window in which lead CM5R and CM5 were correctly detected, while lead CC5 was incorrect. Fig. S2 shows an example of AFl in which only lead CM5R is correctly classified. 

Since AF and AFl are associated with a higher heart rate, AF detectors learn that a higher heart rate is associated with a higher likelihood of AF. This may explain the non-negligible proportion of FPs classified as non-$\textrm{AF}_{\textrm{l}}$ tachycardia. A large proportion of FP and FN windows were associated with poor signal quality (27\% and 26\%, respectively), much higher than the proportion of TP and TN windows (10\% and 10\%, respectively). Among FPs, there was an estimated large proportion of examples with an incorrect label (27.9\%). This reflects the limits in expert manual annotations of AF using Holter ECG. Similar to previous work~\cite{Lewis2011, Ip2020}, we report a large proportion of FP (15.4\%) associated with the presence of ectopic beats.

While RawECGNet demonstrated high generalization performance, the performance on the two external data sets (F1 of 0.91--0.94 and 0.93) were still inferior to the performance reached on the local test set (F1 of 0.95--0.96). This suggests that RawECGNet can be further improved. Training in multiple channels significantly improved the generalization performance. However, the UVAF data set was limited to three channels. Consequently, learning a representation from annotated 12-lead ECGs may help to further improve the generalization performance of RawECGNet by creating a more exhaustive representation of AF across different lead systems. Large 12-lead ECG data sets are open access~\cite{Reyna20221} which may be suitable for this purpose. However, the 12-lead ECGs are typically 10-s in length, and, consequently, the importance of window size on performance will have to be evaluated. ECG foundation models may also be useful in extracting meaningful features and improving generalization~\cite{Viad2023}. Another direction for future work should put more emphasis on improving the discrimination between AF and AFl episodes.

For detectors based on feature engineering, better performance has been reported for rhythm-based features only than that obtained for a combination of rhythm- and morphological features~\cite{SornmoAFbookDetection}. Although this relation may seem surprising, it can be explained by the fact that the performance of detectors of the latter type is more sensitive to noise since an important aim of morphological features is to characterize low-amplitude atrial waves. Moving to the performance of the two deep-learning-based detectors, RawECGNet and ArNet2, the relation is the opposite. The results in Table~\ref{tab:generalization_resutls} show that information on morphology actually matters since RawECGNet performs better than ArNet2 which uses RR~intervals as input. 

The present research proposed and evaluated the new deep learning model RawECGNet for the detection of AF and AFl episodes. RawECGNet was validated across all leads of two external Holter data sets that were manually annotated at the single-beat level. This approach offers robust evidence of RawECGNet's performance. However, considering the potential variations across ethnicities and the quality of ECGs recorded by single-lead wearable systems, it will be crucial to further assess RawECGNet across a variety of ethnic groups and on wearable ECG data sets. Finally, we performed a thorough error analysis which provides further research directions.



\printbibliography






\section{Supplementary}

\supplementarysection

\begin{figure}[H]
\centering
\includegraphics[page=1,width=1\columnwidth,]{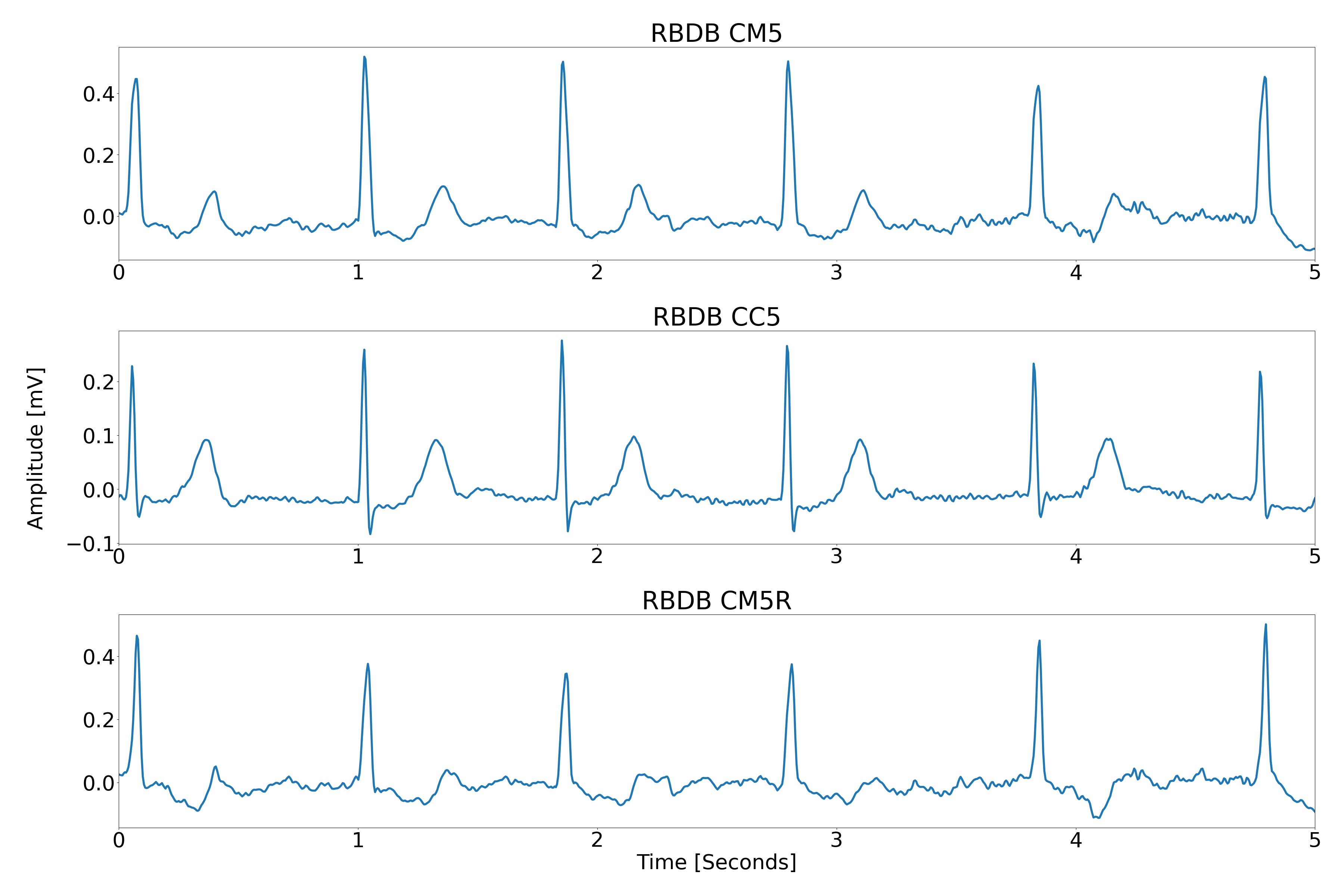}
\caption{Example of an  AF ECG segment form RBDB where RawECGNet correctly classified lead CM5 and lead CM5R but miss classified lead CC5. This example shows how there is more atrial activity captured in the CM5 and CM5R leads than in the CC5 lead.}
\label{RBDB_CC5_FN_example}
\end{figure}

\begin{figure}[H]
\centering
\includegraphics[page=1,width=1\columnwidth,]{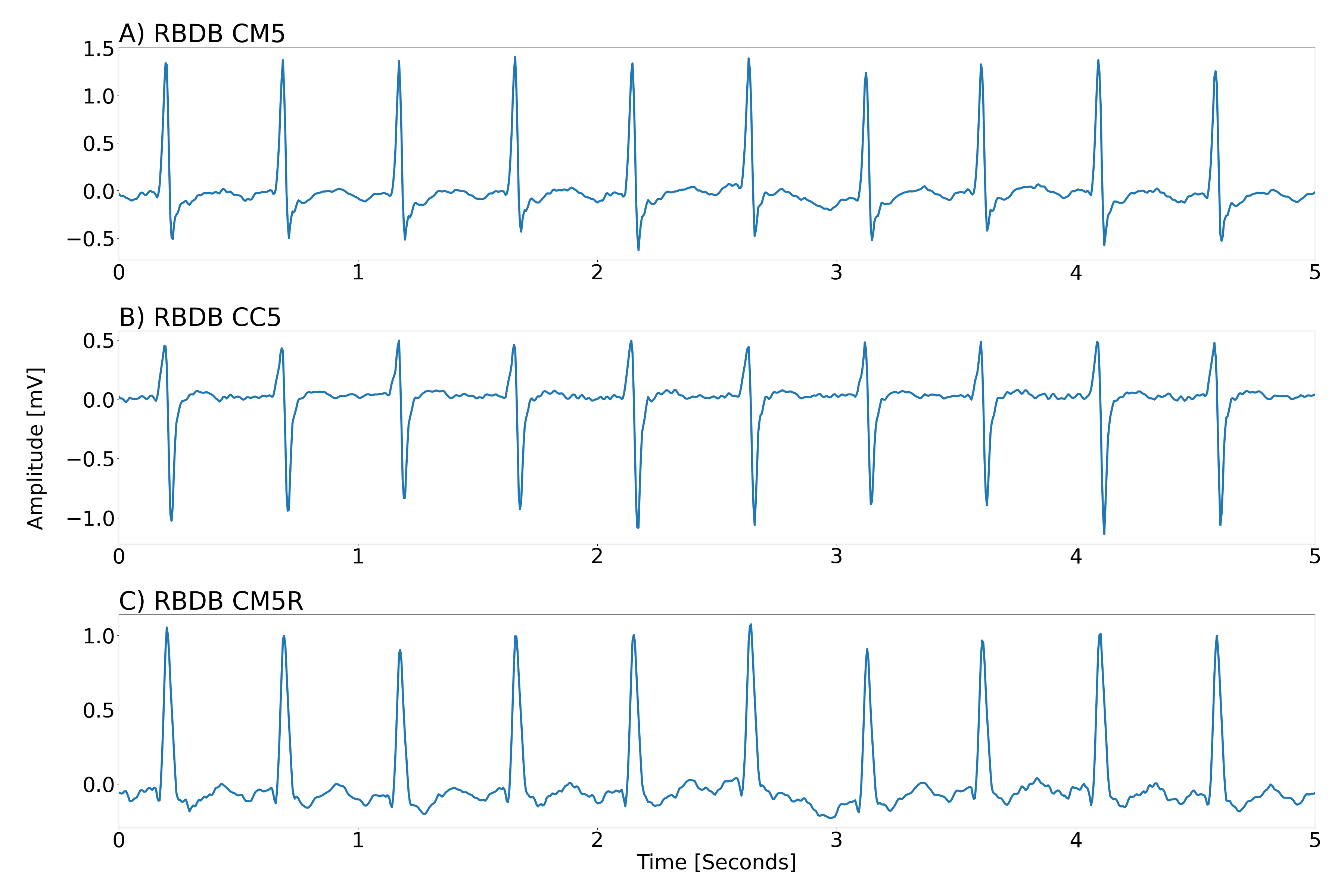}
\caption{Example of an ECG segment form RBDB of an AFl event where RawECGNet correctly classified lead CM5R but miss classified lead CC5 and lead CM5. In CM5R there is the most atrial activity seen}
\label{RBDB_CC5_FN_example_AFL}
\end{figure}

\begin{figure*}[t]
\centering
\includegraphics[page=1,width=1\textwidth,]{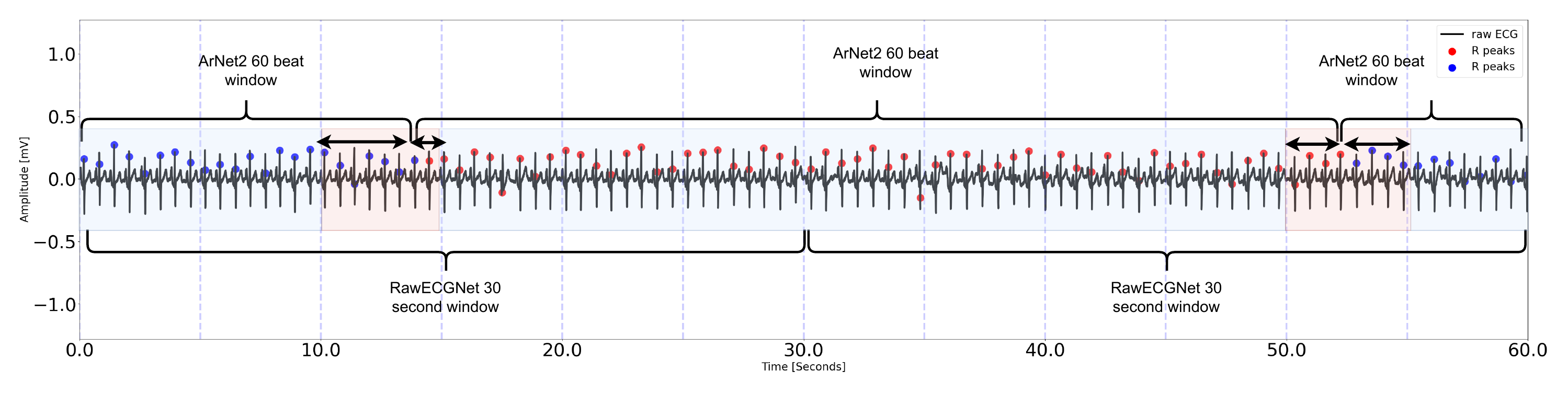}
\caption{Depicted are the 30-s windows RawECGNet uses as input and the 60-beat windows ArNet2 uses as input. For 5-s windows that had more than one window overlapping them, the prediction of the window that contained a greater part of the 5-s window was the prediction used for that window.}
\label{wave_rr_window_comparison}
\end{figure*}


\end{document}